\documentclass[aps,prl,showpacs, twocolumn,floatfix,superscriptaddress,nobalancelastpage]{revtex4-1}
\usepackage{graphicx}
\usepackage{bm}
\usepackage{amsmath}
\usepackage[squaren,Gray]{SIunits} 
\usepackage[caption=false]{subfig}
\usepackage{mathbbol}
\usepackage{xcolor}
\usepackage{subfig}

\providecommand{\abs}[1]{\left\lvert#1\right\rvert}

\providecommand{\moy}[1]{\langle #1 \rangle}
\providecommand{\bra}[1]{\langle #1 \rvert}
\providecommand{\ket}[1]{\lvert #1 \rangle}

\providecommand{\ketbra}[2]{\lvert  #1\rangle \langle #2 \rvert}
\providecommand{\be}{\begin{equation}}
\providecommand{\ee}{\end{equation}}
\providecommand{\ba}{\begin{eqnarray}}
\providecommand{\ea}{\end{eqnarray}}

\newcommand{\One}{\mathbb{1}}
\newcommand{\tr}[1]{\ensuremath{{#1}^{\dagger}}}

\begin{document}

\title{
A general dichotomization procedure to provide qudits entanglement criteria }
\author{Ibrahim Saideh}
\affiliation{Universit\'e Paris\textendash Sud, Institut des Sciences Mol\'eculaires d'Orsay (UMR 8214 CNRS), F-91405 Orsay, France}
\author{Alexandre Dias Ribeiro}
\affiliation{Departamento de F\'{\i}sica, Universidade Federal do Paran\'a, C.P. 19044, 81531-980, Curitiba, PR, Brazil}
\affiliation{Laboratoire Matériaux et Phénomènes Quantiques, Université Denis Diderot, Case 7021, 2 Place Jussieu,
75251 Paris cedex 05, France}
\author{Giulia Ferrini}
\author{Thomas Coudreau}

\author{P\'erola Milman}
\affiliation{Laboratoire Matériaux et Phénomènes Quantiques, Université Denis Diderot, Case 7021, 2 Place Jussieu,
75251 Paris cedex 05, France}
\author{Arne Keller}
\affiliation{Universit\'e Paris\textendash Sud, Institut des Sciences Mol\'eculaires d'Orsay (UMR 8214 CNRS), F-91405 Orsay, France}

\begin{abstract}
We present a general strategy to derive entanglement criteria which consists in performing a mapping from qudits to qubits that preserves the separability of the parties and SU(2) rotational invariance. Consequently, it is possible to apply the well known positive partial transpose criterion to reveal the existence of quantum correlations between qudits. We discuss some examples of entangled states that are detected using the proposed strategy. Finally, we demonstrate, using our scheme,  how some variance based
entanglement witnesses can be generalized from qubits to higher dimensional spin systems.
\end{abstract}

 \vskip2pc 
 
\maketitle
A necessary and sufficient condition to assert the separability of a given general quantum state is an open problem.
For bipartite quantum systems formed by two 2-level systems (qubits) or a 2-level system and a 3-level system (qutrit), the Peres-Horodecki criterion~\cite{Peres,PPT}, that is the positivity of the partial transpose (PPT) of the quantum system density matrix, provides such a condition for separability. We still lack such a criterion to fully characterize separability in higher dimensional  or in multipartite quantum systems.
Although the PPT test constitutes a sufficient condition for detecting bipartite entanglement in the case of systems of arbitrary dimension, it requires not only local manipulation of each party, but also a full reconstruction of the state density matrix. 
Such  requirements may be prohibitive from an experimental perspective. This is why entanglement witnesses~\cite{HorodeckiRev},  which are sufficient criteria for detecting bipartite or even multipartite entanglement based on less demanding measurements, have attracted so much interest lately~\cite{Ex1}.

In this Letter, we present a general scheme to detect entanglement in systems of arbitrary (finite) dimension based on the mapping of  qudits to qubits. The proposed mapping, which actually constitutes a general and operational formulation for dichotomization, preserves the separability of the subsystems, ensuring that it does not create entanglement that did not exist in the original system. Therefore, if the mapped qubits system is entangled, we can assert that the corresponding qudits are also entangled. In this way, the proposed mapping enables the application of entanglement criteria originally derived for qubit systems to qudit ones. 

We start by defining the mapping $\mathcal{M}_U$  of a density operator $\rho$ acting on the $d$-dimensional Hilbert space $\mathcal{H}^{(d)}$  to a density operator $\sigma = \mathcal{M}_U(\rho)$ acting on a 2-dimensional Hilbert space $\mathcal{H}^{(2)}$: 
\begin{align}
\mathcal{M}_U: \mathcal{B}(\mathcal{H}^{(d)}) & \rightarrow \mathcal{B}(\mathcal{H}^{(2)}) \nonumber\\ 
\rho & \rightarrow \mathcal{M}_U(\rho) = \text{Tr}_D
\left[ U\left(\rho\otimes\ket{0}\bra{0}\right)U^{\dagger}\right].
\label{eq:mapdef}
\end{align}
$\ket{0}$ is an arbitrary fiducial state in $\mathcal{H}^{(2)}$, $U$ is a linear operator from  the tensor product space $\mathcal{H}^{(d)}~\otimes~\mathcal{H}^{(2)}$ to another Hilbert space $\mathcal{H}^{(D)}\otimes\mathcal{H}^{(2)}$ with $D \geq d$,
and $\text{Tr}_D$ means a partial trace over the $\mathcal{H}^{(D)}$ part only.  $\mathcal{B}(\mathcal{H})$ denotes the space of the bounded operators acting on $\mathcal{H}$.
To preserve hermiticity, positivity and trace of the density matrix, it is enough to require  that $U$ is an isometry, that is $U^\dagger U = \One_{2d}$ , where $\One_{2d}$ is the identity on  $\mathcal{H}^{(d)}\otimes\mathcal{H}^{(2)}$\footnote{An isometric operator fulfills $U^{\dagger}U=\One_{2d}$, but the additional identity $UU^{\dagger}=\One_{\mathcal{H}^{(D)}\otimes\mathcal{H}^{(2)}}$ is not required as it would be the case if $U$ is a unitary}.

Since we are interested in detecting bipartite entanglement in the state described by $\rho$ acting on $\mathcal{H}^{(d_1)}\otimes \mathcal{H}^{(d_2)}$, we define a mapping
$\mathcal{M}_{U_1U_2}$ as follows:
\begin{align}
\mathcal{M}_{U_1U_2}:\mathcal{B}(\mathcal{H}^{(d_1)}\otimes \mathcal{H}^{(d_2)})& \rightarrow \mathcal{B}(\mathcal{H}^{(2)} \otimes \mathcal{H}^{(2)}) \nonumber\\ 
\rho & \rightarrow r = \mathcal{M}_{U_1U_2}(\rho),
\end{align}
 with:
\be
\mathcal{M}_{U_1U_2}(\rho) 
= \text{Tr}_{D_1D_2}
\left[ U_1\otimes U_2\left(\rho\otimes\ket{0}\bra{0}\otimes\ket{0}\bra{0}\right)U_1^{\dagger}
\otimes U_2^{\dagger}\right],
\ee
where $U_i$ is an isometric operator   acting from $\mathcal{H}^{(d_i)}\otimes \mathcal{H}^{(2)}$ to $\mathcal{H}^{(D_i)}\otimes \mathcal{H}^{(2)}$ $ (\text{for }~i=1,2)$ and $\text{Tr}_{D_1D_2}$ means a partial trace over $\mathcal{H}^{(D_1)}\otimes \mathcal{H}^{(D_2)}$ states.   Analogously to the case with $\mathcal{M}_U$, the mapping $\mathcal{M}_{U_1U_2}$ preserves positivity and trace of the density operator. Furthermore, it also preserves separability: indeed, if $\rho$ is a 2-qudit product state 
$\rho = \rho_1\otimes \rho_2$, then $r=\mathcal{M}_{U_1U_2}(\rho)= \mathcal{M}_{U_1}(\rho_1)\otimes\mathcal{M}_{U_2}(\rho_2)$ is also a product state. This property can be extended to all separable states, pure or mixed, by convexity.
Therefore if $r$ is an entangled state, we are sure that the original 2-qudit state $\rho \in \mathcal{B}(\mathcal{H}^{(d_1)}\otimes \mathcal{H}^{(d_2)})$ is entangled. As all entangled states in $ \mathcal{B}(\mathcal{H}^{(2)} \otimes \mathcal{H}^{(2)})$ have one nonpositive eigenvalue for the partially transposed matrix, two natural questions arise: given an entangled density matrix $r$, what type of entanglement is  present in the original  2-qudit state defined in $\mathcal{B}(\mathcal{H}^{(d_1)}\otimes \mathcal{H}^{(d_2)})$ ? And  how does the set of detectable entangled states depend on the isometries $U_1$ and $U_2$ that are used to implement the mapping? 

In the present Letter, we provide some answers to these questions in the operationally relevant case where  the isometries $U_{1,2}$ correspond to a mapping which preserves the  SU(2) rotational invariance and, in addition,  lead to an entanglement witness which can be easily implemented experimentally.

We start by remarking that the isometry $U$  which characterizes the mapping $\mathcal{M}_U$ can be parametrized by 4 linear  operators $A_i \, (i=0,1,2,3)$ from $\mathcal{H}^{(d)}$ to $\mathcal{H}^{(D)}$
 as follows:
\be
\label{eq:UA}
U = \sum_{i=0}^3 A_i \otimes \sigma_i,
\ee
where $\sigma_i$'s, for $i=1,2,3$, are the Pauli matrices and $\sigma_0 = \One$ is the identity operator in $\mathcal{H}^{(2)}$. The isometric property $U^{\dagger}U=\One$ implies that:
\begin{align}
&\tr{\vec{A}}\cdot\vec{A}+\tr{A_0}A_0=\One,\nonumber \\
&\tr{\vec{A}}A_0+\tr{A_0}\vec{A}+i\tr{\vec{A}}\wedge\vec{A}=\vec{0}.
\label{eq:AA}
\end{align} 
With the parametrization Eq.~\eqref{eq:UA} the mapping Eq.~\eqref{eq:mapdef} can be written as:
\be
\label{MM}
\mathcal{M}_U\left(\rho\right)=\sum_{i,j=0}^3\moy{\tr{A_j}A_i}_\rho\,\sigma_i\ketbra{0}{0}\sigma_j.
\ee

We proceed by introducing the following  intuitive and convenient specific case for this mapping:
\be
\mathcal{M}_U(\rho) = \frac{1}{2}\left(\One+ \frac{1}{j}\sum_{i=1}^{3}\moy{J_i}_\rho \sigma_i\right),
\label{eq:ourMapU}
\ee
where $J_i \,  (i=1,2,3)$ are the 3 Cartesian angular momentum components. They are the generators of SU(2) rotations around $x,~y,$ and $z$ axis, in the $d$-dimensional Hilbert space $\mathcal{H}^{(d)}$. The denominator $j$ is the largest eigenvalue of $J_i$, i.e.,  it is such that $d=2j+1$. Finally, $\moy{J_i}_{\rho}=\text{Tr}[\rho J_i]$ denotes the expectation value of $J_i$.

The mapping $\mathcal{M}_U$ given by Eq.~\eqref{eq:ourMapU} can be shown to be  a valid mapping defined from Eq.~\eqref{eq:mapdef}, by exhibiting the corresponding  $A_i$ operators of Eq.~\eqref{eq:UA}. We note that the $A_i$ operators can be expressed in a simple form with the help of the two bosonic annihilation  operators $a$ and $b$ corresponding to the 
Schwinger representation:  $J_+ = a^{\dagger}b$, $ J_- =b^{\dagger}a$, where 
$J_{\pm} = J_1 \pm iJ_2$, and $ J_3= \frac{1}{2}(a^{\dagger}a - b^{\dagger}b)$ with the restriction $a^{\dagger}a+b^{\dagger}b=2j$.
By a straightforward calculation the operators $A_i$ realizing the isometry $U$ through Eqs.~\eqref{eq:UA} and Eqs.~\eqref{eq:AA} can be shown to be~\cite{supp}:
\begin{align}
A_0 &= \frac{1}{2\sqrt{2}}\left(\frac{a}{\sqrt{j}} + \frac{a^{\dagger}}{\sqrt{j+1}}\right), 
A_3= \frac{1}{2\sqrt{2}}\left(\frac{a}{\sqrt{j}} - \frac{a^{\dagger}}{\sqrt{j+1}}\right), \nonumber \\ 
A_1 &= \frac{1}{2\sqrt{2}}\left(\frac{b}{\sqrt{j}} - \frac{b^{\dagger}}{\sqrt{j+1}}\right),
iA_2 = \frac{1}{2\sqrt{2}}\left(\frac{b}{\sqrt{j}} + \frac{b^{\dagger}}{\sqrt{j+1}}\right).
\label{eq:Aschwinger}
\end{align}
We emphasize that this  mapping is practical, as the 3 expectation values $\moy{J_i}_{\rho}$ can be easily measured. In addition, it conserves the rotational invariance. Indeed, suppose that we perform a rotation $\mathcal{R}_{\vec{n}}(\alpha)$ by an angle $\alpha$ around a given vector $\vec{n}$. Then,  $\rho$ is transformed as
$\rho' = e^{-i\alpha\vec{J}\cdot\vec{n}} \rho  e^{i\alpha\vec{J}\cdot\vec{n}}$. It is not difficult to show that $\rho'$ is mapped to the rotated qubit~:
\be
\mathcal{M}_U( e^{-i\alpha\vec{J}\cdot\vec{n}} \rho  e^{i\alpha\vec{J}\cdot\vec{n}}) = e^{-i\alpha\vec{\sigma}\cdot\vec{n}/2} \mathcal{M}_U(\rho)  e^{i\alpha\vec{\sigma}\cdot\vec{n}/2}.
\label{eq:rotInv}
\ee
The invariance displayed in Eq.~\eqref{eq:rotInv} is a consequence of the simple  vectorial relations:
\be
\moy{\mathcal{R}^{-1}_{\vec{n}}(\alpha)\left[ \vec{J}\right]}_{\rho}\cdot \vec{\sigma} = \moy{\vec{J}}_{\rho'}\cdot\sigma = \moy{\vec{J}}\cdot \mathcal{R}_{\vec{n}}(\alpha)\left[ \vec{\sigma}\right],
\ee
where $\vec{J}$ is the vector whose 3 components are the 3 operators $J_i$, for $i=1,2,3$, and  $\mathcal{R}^{-1}_{\vec{n}}(\alpha)\left[ \vec{J}\right]$ is the corresponding vector in the rotated frame. Since SO(3) and SU(2) are isomorphic, we can apply all possible unitaries to the mapped qubit by rotating the original angular momentum correspondingly.

Now, we consider the case of a 2-qudit state $\rho$ in $ \mathcal{B}(\mathcal{H}^{(d_1)}\otimes \mathcal{H}^{(d_2)})$ and use the mapping $\mathcal{M}_{U_1U_2}$, where each $U_i (i=1,2)$ implements a mapping as the one given by Eq.~\eqref{eq:ourMapU}. The resulting mapping $\mathcal{M}_{U_1U_2}$ can then be written explicitly as follows:
\begin{align}
\label{eq:nondiagmap}
\begin{split}
\mathcal{M}_{U_1U_2}(\rho)  &= \frac{1}{4}\left[ \One + \frac{1}{j^{(d_1)}}\sum_{i=1}^3\moy{J^{(d_1)}_i\otimes \One}_{\rho} \sigma_i\otimes\One \right.\\ 
&+  \frac{1}{j^{(d_2)}}\sum_{i=1}^3 \moy{\One\otimes J^{(d_2)}_i}_{\rho}\One \otimes \sigma_i +  \\
&+ \left. 
 \frac{1}{j^{(d_1)}j^{(d_2)}}\sum_{i,j=1}^3 \moy{J_i^{(d_1)}\otimes J_j^{(d_2)}}_{\rho}\sigma_i\otimes\sigma_j
\right],
\end{split}
\end{align}
where $J_i^{(d_k)}$ is the $i$-th  $(i=1,2,3)$ angular momentum component  on the $d_k$-dimensional Hilbert space $\mathcal{H}^{(d_k)}$  $(k=1,2)$, with $d_k = 2j^{(d_k)} +1$. 

An  important property of this particular mapping is that the partial transpose (PT) of a 2-qudit state is mapped to the PT of its corresponding 2-qubit state, i.e., $\mathcal{M}_{U_1U_2}(\rho^{T_2})=\mathcal{M}_{U_1U_2}(\rho)^{T_2}$, where $^{T_2}$ is the PT with respect to the second party. This follows directly from the fact $\text{Tr}[A^{T_2}B]=\text{Tr}[AB^{T_2}]$ for $A$ and $B$ acting on  $\mathcal{H}^{(d_1)}\otimes \mathcal{H}^{(d_2)}$ and that  the transpose of the components of the angular momentum operator fulfills  $J_1^T = J_1$, $J_2^T=-J_2$, and $J_3^T=J_3$. 
As a result, taking into account positivity preservation, a 2-qudit  state which remains positive after a partial tranpose (a PPT state) is mapped to a PPT 2-qubit state
\footnote{We remark that our criterion does not allow the detection of bound entanglement, as it preserves the negativity of the PT}.
Moreover, it leads to an operationally easy to implement entanglement witness based on second order correlations.  This is achieved by using the following substitutions:
\ba
\begin{split}
\label{rTdef}
\moy{\sigma_i\otimes\One}_{\mathcal{M}_{U_1U_2}(\rho)}=\dfrac{\moy{J^{(d_1)}_i\otimes \One}_{\rho}}{j^{(d_1)}}&\equiv & r^1_i,\\ 
\moy{\One\otimes\sigma_i}_{\mathcal{M}_{U_1U_2}(\rho)}=\dfrac{\moy{\One\otimes J^{(d_2)}_i}_{\rho}}{j^{(d_2)}}&\equiv & r^2_i,\\ 
\moy{\sigma_i\otimes \sigma_j}_{\mathcal{M}_{U_1U_2}(\rho)}=\dfrac{\moy{J_i^{(d_1)}\otimes J_j^{(d_2)}}_{\rho}}{j^{(d_1)}j^{(d_2)}}&\equiv & T_{ij},
\end{split}
\ea
 which are  direct consequences of Eq.~\eqref{eq:nondiagmap}.
Therefore we consider the following form for the general mapped state:
\be
\label{mappingform}
\mathcal{M}_{U_1U_2}(\rho)=\frac{1}{4}\left[\One +\vec{r}^1.\vec{\sigma}\otimes\One +\vec{r}^2.\One\otimes\vec{\sigma}+\sum_{i=1}^3 T_{ii}\sigma_i\otimes\sigma_i\right],
\ee
where we assume that the rotations needed to diagonalize $T_{ij}$ have been performed. We have introduced the vectors $\vec{r}^j$ ($j=1,2$), which have as components the $r_i^j$ ($i=1,2,3$) defined in Eq.~\eqref{rTdef}, after the mentioned needed rotations.


Once we have the form of Eq.~\eqref{eq:nondiagmap} or Eq.~\eqref{mappingform}, the best entanglement witness that we can use is the  Peres-Horodecki criterion~\cite{PPT} relying on the positiveness of the partial transpose.
 This criterion can be simplified by considering the  geometric picture which was developed in Ref.~\cite{Horodecki1996}.
 In this paper, it was shown that for the states of the form given by Eq.~\eqref{mappingform}, the vector $\vec{T}=\lbrace T_{11},T_{22},T_{33}\rbrace \in \mathbb{R}^3$ must lie within a tetrahedron with vertices $\lbrace(-1,-1,-1),(-1,1,1),(1,-1,1),(1,1,-1)\rbrace$, to fulfill the positiveness requirement. Each of the  4 vertices of this tetrahedron is  reached when the 2-qubit state is one of the  four Bell-states $\ket{\Phi^{\pm}}=\frac{\ket{00}\pm\ket{11}}{\sqrt{2}}$, $\ket{\Psi^{\pm}}=\frac{\ket{01}\pm\ket{10}}{\sqrt{2}}$. In this picture, the separable states are those for which the vector  $\vec{T}=\lbrace T_{11},T_{22},T_{33}\rbrace \in \mathbb{R}^3$ lies within the octahedron with vertices    $\lbrace(\pm 1,0,0),(0,\pm 1,0),(0,0,\pm 1)\rbrace$. This last property can be put in the following  more compact form: 
for any separable state of the form given by  Eq.~\eqref{mappingform}, $\vec{T}$ verifies the inequality:
\be
\abs{T_{11}}+\abs{T_{22}}+\abs{T_{33}}\leq 1. 
\label{Ts}
\ee   
 Using the definitions given by Eq.~\eqref{rTdef}, Eq.~\eqref{Ts} can be re-expressed as a 2-qudit entanglement criterion: \\
{\it For any separable state $\rho$ acting on $\mathcal{H}^{(d_1)}\otimes \mathcal{H}^{(d_2)}$, the vector $\lbrace \moy{J_i^{(d_1)}\otimes J_i^{(d_2)}}_{\rho} :i=1,2,3\rbrace$ verifies the following inequality:}
\begin{align}
\label{crit}
\abs{\moy{J_1^{(d_1)}\otimes J_1^{(d_2)}}_{\rho}}+&\abs{\moy{J_2^{(d_1)}\otimes J_2^{(d_2)}}_{\rho}} +\nonumber \\
+&\abs{\moy{J_3^{(d_1)}\otimes J_3^{(d_2)}}_{\rho}}\leq {j^{(d_1)}j^{(d_2)}}.
\end{align}
Therefore, all states that violate inequality~\eqref{crit} lie  outside the octahedron and are thus entangled.
This 2-qudit entanglement criterion has the advantage of being very simple to test experimentally. 
One can notice that there are no 2-qudit states that are mapped to any of the 2-qubit Bell states~\cite{supp}. Therefore, the vertices of the tetrahedron do not belong to the image of our mapping.






To have an insight about the efficiency  of our criterion to detect entanglement,  we apply it to  two known families of qudit states that have been extensively studied  in Refs.~\cite{B2006,B2008,Be2008,B2009}. 
 From now on, we suppose that the two qudits are of the same dimension, that is  $d_1=d_2\equiv d=2j+1$. First, we recall the family of $d^2$ maximally entangled 2-qudit states $\ket{\Omega_{kl}} (k,l=0,1,\cdots, d-1)$ that  generalize the  four  2-qubit Bell states~\cite{B2006,B2008,Be2008,B2009}:
\ba
\label{Bell}
\hspace{-1.4em}\ket{\Omega_{kl}}= W_{kl}\otimes\One\ket{\Omega_{00}},~\text{with}~  \ket{\Omega_{00}}=\frac{1}{\sqrt{d}}\sum_{m=0}^{d-1}\ket{m,m},
\ea
where the $d^2$ $W_{kl}$ operators acting on the first qudit are the Weyl operators defined as
\ba
 W_{kl}\ket{m}=e^{\frac{i2\pi k(m-l)}{d}}\ket{(m-l)_{\text{mod }d}}.
\ea
These 2-qudit Bell states $P_{kl}=\ketbra{\Omega_{kl}}{\Omega_{kl}}$ are locally maximally mixed states, that is, by taking their partial trace one obtains the maximally mixed state $\frac{\One}{d}$. It is interesting to notice that they are mapped by Eq.~\eqref{mappingform} to a locally maximally mixed 2-qubit state. Indeed, we have:
\begin{equation}
\vec{r}_{kl}^1=\dfrac{\moy{\vec{J}\otimes \One}_{P_{kl}}}{j}=\vec{0}\quad\text{and}\quad \vec{r}_{kl}^2=\dfrac{\moy{\One\otimes \vec{J}}_{P_{kl}}}{j}=\vec{0}. \nonumber
\end{equation}
For such states, our simple criterion Eq.~\eqref{crit} is as strong as the PPT criterion applied to states given by Eq.~\eqref{mappingform} and detects all entangled locally maximally mixed states that are detected by the latter.
We can thus say that every maximally entangled 2-qudit state $\ket{\Omega_{kl}}$ is detected by our criterion Eq.~\eqref{crit}, even though these states are not mapped to 2-qubit Bell states~\cite{supp}. Instead, they are mapped to mixed states that are locally maximally mixed, that is, a convex sum (statistical mixture) of 2-qubit Bell states.

We proceed by exploring the statistical mixtures of maximally entangled 2-qudit pure states that can be detected by our criterion  Eq.~\eqref{crit}. We start by applying our criterion to the so called Werner states which form a good description of the effects of phase and depolarizing noise in maximally entangled states~\cite{Horodecki1997,B2008}:
 \be
 \rho_{\alpha}=\alpha\ketbra{\Omega_{00}}{\Omega_{00}}+\dfrac{1-\alpha}{d^2}\One,~\text{with }~\frac{-1}{d^2-1}\leq\alpha\leq 1.
 \ee
The bounds for the parameter $\alpha$ are such that  $\rho_{\alpha}$ is positive. It is known~\cite{Horodecki1997} that $\rho_{\alpha}$ is entangled iff $\frac{1}{d+1}<\alpha\leq 1$. A straightforward application of our criterion Eq.~\eqref{crit} to $\rho_\alpha$ gives that if $\alpha j\left(j+1\right) > j^2$ then $\rho_\alpha$ is entangled.  Therefore, our criterion can detect all the entangled states $\rho_{\alpha}$ for $\alpha \in [\frac{j}{\left(j+1\right)},1]$. Recalling that $d=2j+1$, we realize that entangled states with $\frac{1}{2\left(j+1\right)}<\alpha\leq \frac{j}{\left(j+1\right)}$ are not detected.

As a more specific example, we now consider  the  3-parameter family of 2-qudit states:
\ba
\label{rhod_abc}
\rho_{\alpha,\beta,\gamma}&=&\dfrac{1-\alpha-\beta-\gamma}{(2j+1)^2}\One+\alpha P_{00}\nonumber \\&+&\dfrac{\beta}{2j}\sum_{i=1}^{2j}P_{i0}+\dfrac{\gamma}{2j+1}\sum_{i=0}^{2j}P_{i1},
\ea
where the $P_{kl}$ are the projectors on the $\ket{\Omega_{kl}}$ states. 
This family of states is a generalization to  arbitrary dimensional qudits  of the 2-qutrit states originally  introduced in Refs.~\cite{B2008,Be2008,B2009} to study bound entanglement. Density matrices as given 
in Eq.~\eqref{rhod_abc}  are interesting because their eigenvalues and those of their partial transpose can be explicitly expressed as a function of the parameters $\alpha$, $\beta$ and $\gamma$~\cite{supp}. This allows to locate the set of PPT $\rho_{\alpha,\beta,\gamma}$  in the space spanned by  $\alpha,\beta,\text{and}~\gamma$.
To ensure positivity, the parameters $\alpha,\beta,\text{and}~\gamma$, must verify the following inequalities~\cite{supp}:
\begin{eqnarray}
 & \dfrac{1-\alpha-\beta-\gamma }{(2j+1)^2} \geq 0  & , \alpha+\dfrac{1-\alpha-\beta-\gamma }{(2j+1)^2}  \geq 0 \nonumber \\ 
& \dfrac{\beta}{2j}+\dfrac{1-\alpha-\beta-\gamma }{(2j+1)^2} \geq 0 & ,  \dfrac{1-\alpha-\beta+2j\gamma }{(2j+1)^2} \geq 0. \nonumber
\end{eqnarray}
These inequalities define the interior of a tetrahedron. Now, applying our criterion Eq.~\eqref{crit},
 we obtain that all separable states $\rho_{\alpha,\beta,\gamma}$ given by Eq. (19) are such that:
\begin{equation}
\label{critabc}
\dfrac{\abs{ (j+1) (\alpha+\beta) +(j-2) \gamma }+(j+1)  \abs{2 \alpha-\beta/j} }{3j}\leq 1.
\end{equation}
All states $\rho_{\alpha,\beta,\gamma}$ for which the inequality Eq~\eqref{critabc} is violated are entangled.

In order to have an idea about the efficiency of criterion Eq.~\eqref{crit}, we have calculated the eigenvalues of the PT of $\rho_{\alpha,\beta,\gamma}$ and thus obtained explicit conditions on $\alpha, \beta,\text{and}~\gamma$ for $\rho_{\alpha,\beta,\gamma}^{T_2}$ to be positive~\cite{supp}.
\begin{figure}[ht]
\hspace{0cm}(a)~\includegraphics[width=0.22\textwidth]{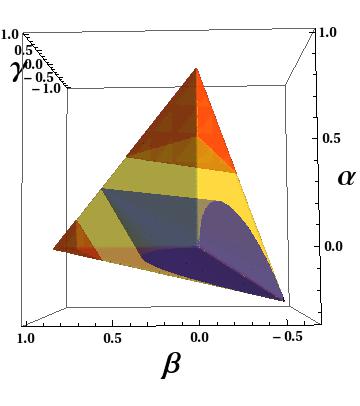}~(b)~\includegraphics[width=0.22\textwidth]{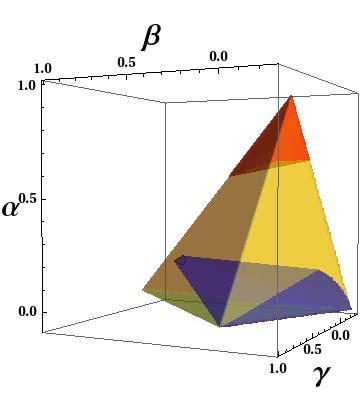}
\caption{\label{fig:F1} Geometrical representation of states $\rho_{\alpha,\beta,\gamma}$~[Eq.~\eqref{rhod_abc}] in parameter space for (a)~$2j+1=3$ and (b) $2j+1=5$. All physical states lie inside the tetrahedron whereas the PPT states lie inside the blue region [a cone for the 2-qutrit case (a)]. Red regions depict entangled states detected using our criterion Eq.~\eqref{crit}, whereas the yellow region hosts the non detected entangled states by this criterion. }
\end{figure}%
In Fig.~\ref{fig:F1}, we present the tetrahedron of positive states $\rho_{\alpha,\beta,\gamma}$ given by Eq.~\eqref{rhod_abc} in parameter space for the cases $j=1$ (Fig.~\ref{fig:F1}.a) and $j=2$ (Fig.~\ref{fig:F1}.b). 
The set of PPT  states are depicted in blue, so  the remaining volume of the tetrahedron corresponds to entangled states. We have represented in red the states that are detected by our criterion Eq.~\eqref{crit}. We see clearly that it detects a significative part of the entangled states parametrized by Eq.~\eqref{rhod_abc}. Nevertheless, comparing Fig.~\ref{fig:F1}.a and Fig.~\ref{fig:F1}.b, we note that this volume decreases when $j$ increases from $j=1$ to $j=2$.
However, different criteria which present a better scaling with dimension for this particular family of states can be found  by  changing the isometry $U$ in~Eq.~\eqref{eq:UA} (or equivalently changing the corresponding $A_i$ operators in Eq.~\eqref{eq:Aschwinger}, used to map each qudit to a qubit).

%

Until now, we only have considered the entanglement of two qudits. Now, we adress the problem of detecting entanglement in a large $N$ qudits system.
An interesting consequence of our mapping is that it can be easily extended to map a system of $N$ qudits to a system of $N$ qubits.
Indeed, by applying Eq.~\eqref{eq:ourMapU} individually to each qudit, the separability among the parties is preserved. If we denote $L_\alpha=\frac{J_\alpha}{j} (\alpha=1,2,3)$ and $L_0=\One$, then the  $N$-qubit mapped state corresponding to  the $N$-qudit state $\rho$ can be written as
\begin{eqnarray}
\mathcal{M}_{\otimes^{N}U} (\rho)=\dfrac{1}{2^N}\sum_{\vec{k}\in \lbrace 0,1,2,3\rbrace ^N}\moy{\otimes_{i=1}^N L_{k_i}}_{\rho}\otimes_{i=1}^N \sigma_{k_i}.
\end{eqnarray}
An immediate consequence is that for any $\vec{k} \in \lbrace 0,1,2,3\rbrace^N$, we have 
\be
\moy{\otimes_{i=1}^N L_{k_i}}_{\rho}=\moy{\otimes_{i=1}^N \sigma_{k_i}}_{\mathcal{M}_{\otimes^{N}U} (\rho)}.
\label{eq:substitution}
\ee
By using this property to compute first and second order correlations, we can provide an alternative derivation of the spin squeezing inequalities detecting $N$-qudit entanglement from the $N$-qubit ones introduced in Ref.~\cite{Toth}. 
It was shown in Ref.~\cite{Toth2007} that all separable $N$-qubit states satisfy the following inequalities:
\begin{align}
\begin{split}
&
\langle\hat{\mathcal{S}}_1^2 \rangle+\langle\hat{\mathcal{S}}_2^2 \rangle +
\langle\hat{\mathcal{S}}_3^2 \rangle  \le \frac{N\left(N+2\right)}{4} ,\\&
(\Delta\hat{\mathcal{S}}_1)^2 +(\Delta\hat{\mathcal{S}}_2)^2 
+(\Delta\hat{\mathcal{S}}_3)^2  \ge \frac{N}{2} , \\&
\langle\hat{\mathcal S}^{\mathrm{lsq}}_\alpha  \rangle +
\langle\hat{\mathcal S}^{\mathrm{lsq}}_\beta  \rangle - (N-1)(\widetilde\Delta\hat{\mathcal{S}}_\gamma)^2
\le \frac{N(N-1)}{4} ,\\& 
\langle \hat{\mathcal S}_\alpha^{\mathrm{lsq}} \rangle
-(N-1)[(\widetilde\Delta\hat{\mathcal{S}}_\beta)^2 +(\widetilde\Delta\hat{\mathcal{S}}_\gamma)^2]
\le \frac{N(N-1)}{4}.
\end{split}
\label{in}
\end{align}
where $\hat{\mathcal S}_\alpha=\frac{1}{2}\sum_{i=1}^N \sigma_{\alpha}^i$ is the collective spin operator in direction $\alpha$. The indexes
$(\alpha,\beta,\gamma)$ may assume any permutation of $(1,2,3)$ and the following definitions have been used:
\begin{equation}
\begin{array}{lll}
(\Delta\hat{\mathcal{S}}_\alpha)^2 &\equiv& 
\langle \hat{\mathcal{S}}_\alpha^2\rangle -\langle \hat{\mathcal{S}}_\alpha \rangle^2,\\
\langle\hat{\mathcal S}^{\mathrm{lsq}}_\alpha  \rangle  &\equiv&
\langle \frac{1}{4}\sum_{i\neq j=1}^N \sigma^i_{\alpha}\otimes\sigma^j_{\alpha} \rangle,\\
(\widetilde\Delta\hat{\mathcal{S}}_\alpha)^2 &\equiv& 
\langle\hat{\mathcal S}^{\mathrm{lsq}}_\alpha  \rangle  -
\langle \hat{\mathcal{S}}_\alpha \rangle^2.
\end{array}
\end{equation}
Using Eq.~\eqref{eq:substitution}, we obtain, starting from the $N$-qubit inequalities~Eqs.~\eqref{in}, the inequalities satisfied by all separable $N$-qudit states ($N$ spins $j$). This is achieved with the following substitutions~\cite{Toth}: 
\begin{eqnarray}
\hat{\mathcal S}_\alpha\rightarrow\frac{1}{2{j}} \hat{\mathcal J}_\alpha,\quad
\hat{\mathcal S}^{\mathrm{lsq}}_\alpha\rightarrow\frac{1}{4{j}^2}\hat{\mathcal J}^{\mathrm{lsq}}_\alpha.
\end{eqnarray}
where $\hat{\mathcal J}_\alpha = \sum_{i=1}^N J_{\alpha}^i$ and
$\hat{\mathcal J}^{\mathrm{lsq}}_\alpha  =  \frac{1}{4}\sum_{i\neq j=1}^N J^i_{\alpha}\otimes J^j_{\alpha}$. Therefore, the relation between the entanglement criterion for $N$-qubit and $N$-qudit systems, which was already considered in  Ref.~\cite{Toth}, can be thought as a simple consequence of the particular mapping explicitly given  by Eqs.~\eqref{eq:Aschwinger} or equivalently by Eq.~\eqref{eq:ourMapU}. 
As a consequence, we have shown that the qudit entanglement revealed by the qudit spin squeezing inequalities can always be recast as qubit-like, or dichotomic, spin squeezing. Thus, qudit spin squeezing inequalities do not evidence entanglement of higher dimension than the qubit squeezing ones. Finally, we notice that by choosing different $A_i$ operators in Eq.~\eqref{eq:UA} we can  expect to find new multipartite qudits entanglement criteria. 

In conclusion, we have presented a general scheme to map qudits to qubits that can be used to define criteria to detect some type of entanglement between qudits. We have applied this general scheme to provide a specific entanglement criterion based on the measurement of qudit--qudit correlations. 
 
 In addition, our results provide a way to classify multi-partite qudit entanglement according to its detectability through dichotomization.  Finally, it opens the interesting question of what are the specific entanglement  types, if any other than bound entanglement, that fail  to be detected by our method.

\acknowledgments{ADR would like to acknowledge financial support from the brazilian agencies CAPES and CNPq via the projects {\it Ci\^encias sem Fronteiras} and {\it Instituto Nacional de Ci\^encia e Tecnologia de Informa\c{c}\~ao Qu\^antica}.}
\bibliography{mapqdit}

\section{Supplemental material}
This supplementary material contains 3 parts: In the first part, we
prove Eq.~(8) of the  main text. In the second part we prove that
 the 2-qubit Bell states are not in the image of our 2-qudit mapping
given by Eq.~(11). Finally, in the third part we introduce the
three-parameter family of 2-qudit states that generalizes the family of
2-qutrit states defined in Ref.~[8]. We show  how we obtain the
separability criterion Eq.~(20) of the main text  as a function of the three parameters defining these states. Next, we calculate the eigenvalues of these states and their partial transpose.
\section{Proof of Eq.~(8)}
In accordance with our general scheme for mapping a qudit state to a
qubit one defined in the text, we need to find operators $A_i$ defining the isometry $U = \sum_{i=0}^3 A_i \otimes \sigma_i$ such that
\be
\label{ourmapexplicit}
\mathcal{M}_U\left(\rho\right)=\sum_{i,j=0}^3\moy{\tr{A_j}A_i}_\rho\,\sigma_i\ketbra{0}{0}\sigma_j 
\ee 
is equal to~Eq.~(7) of main text which we recall here:
\be
\mathcal{M}_U(\rho) = \frac{1}{2}\left(\One+ \frac{1}{j}\sum_{i=1}^{3}\moy{J_i}_\rho \sigma_i\right).
\label{ourmap} 
\ee
 In order for $U$ to be an isometry $U^{\dagger}U=\One$, we have the following conditions:
\begin{eqnarray}
\label{Isometryconditions}
\begin{split}
 A_0^{\dagger}A_0+A_1^{\dagger}A_1+A_2^{\dagger}A_2+A_3^{\dagger}A_3&=&\One,\\
 A_1^{\dagger}A_0+A_0^{\dagger}A_1+iA_2^{\dagger}A_3-iA_3^{\dagger}A_2&=&0,\\
A_2^{\dagger}A_0+A_0^{\dagger}A_2-iA_1^{\dagger}A_3+iA_3^{\dagger}A_1&=&0,\\
A_3^{\dagger}A_0+A_0^{\dagger}A_3+iA_1^{\dagger}A_2-iA_2^{\dagger}A_1&=&0,
\end{split}
 \end{eqnarray}
wich are equivalent to Eq.~(5) of main text.
 Comparing Eq.~\eqref{ourmap} and Eq.~\eqref{ourmapexplicit}, we get \begin{eqnarray}\begin{split}\label{Aconditions}
\hspace{-1em} &-iA_2^{\dagger}A_3+iA_3^{\dagger}A_2+\dfrac{A_1^{\dagger}A_3+A_3^{\dagger}A_1}{2}+\dfrac{iA_0^{\dagger}A_2-iA_2^{\dagger}A_0}{2}=\dfrac{J_1}{2j},&\\
 \hspace{-1em}& iA_1^{\dagger}A_3-iA_3^{\dagger}A_1+\dfrac{A_2^{\dagger}A_3+A_3^{\dagger}A_2}{2}+\dfrac{-iA_0^{\dagger}A_1+iA_1^{\dagger}A_0}{2}=\dfrac{J_2}{2j},&\\
\hspace{-1em}  & -iA_1^{\dagger}A_2+iA_2^{\dagger}A_1+A_3^{\dagger}A_3+A_0^{\dagger}A_0-\dfrac{\mathbb{1}}{2}=\dfrac{J_3}{2j}.&
   \end{split}
 \end{eqnarray}
If we define
$\mathcal{A}_{03}\hspace{-0.2em}=\hspace{-0.2em}\sqrt{j}\left(A_0+A_3\right)$
and
$\mathcal{A}_{12}\hspace{-0.2em}=\hspace{-0.2em}\sqrt{j}\left(A_1+iA_2\right)$,
the set of equations Eqs.~\eqref{Aconditions} can be simplified using
the set of conditions Eqs.~\eqref{Isometryconditions} to
\begin{equation}
\mathcal{A}_{03}^{\dagger}\mathcal{A}_{03}=\dfrac{j+J_3}{2}~,~\mathcal{A}_{12}^{\dagger}\mathcal{A}_{12}=\dfrac{j-J_3}{2}~,~\mathcal{A}_{03}^{\dagger}\mathcal{A}_{12}=\dfrac{J_1+iJ_2}{2}.\nonumber
\end{equation}
These equations can be simplified  with the help of the two bosonic annihilation  operators $a$ and $b$ corresponding to the 
Schwinger representation of the qudit;  $J_+ = a^{\dagger}b$, $ J_- =b^{\dagger}a$ and $ J_3= \frac{1}{2}(a^{\dagger}a - b^{\dagger}b)$ with the restriction $a^{\dagger}a+b^{\dagger}b=2j$ where 
$J_{\pm} = J_1 \pm iJ_2$. We finally get:
\begin{equation}
\mathcal{A}_{03}^{\dagger}\mathcal{A}_{03}=\dfrac{a^{\dagger}a}{2}~,~\mathcal{A}_{12}^{\dagger}\mathcal{A}_{12}=\dfrac{b^{\dagger}b}{2}~,~\mathcal{A}_{03}^{\dagger}\mathcal{A}_{12}
=\dfrac{a^{\dagger}b}{2}.\nonumber
\label{eq:Aconditions}
\end{equation}
It is straightforward to show that the operators $A_i$ defined by
Eq.~(8) of the main text verify the conditions of
Eqs.~\eqref{Isometryconditions} and Eqs.~\eqref{Aconditions}.
Thus, we have proved the validity of mapping~\eqref{ourmap}. We note that the requirement for  $U$ to be an isometry and not a unitary operator is crucial in the present case. Indeed, the operators $A_i$  do not fulfill the conditions for $U$ to be unitary. 

\section{Bell states are not reached}
We proceed to show that none of the four 2-qubit Bell states belongs
to the image of the mapping given by Eq.~(11) of the main text. This can be proved by contradiction:  consider that there exists a state $\rho$ such that $\mathcal{M}_{U_1U_2}(\rho)$ is one of the Bell states, say $\ketbra{\Psi^+}{\Psi^+}$. For this state we have: \ba &&\text{Tr}\left[\ketbra{\Psi^+}{\Psi^+}\sigma_3\right]=~1,\nonumber \label{T33}\\&& \text{Tr}\left[\ketbra{\Psi^+}{\Psi^+}\sigma_2\right]=-1,~ \text{Tr}\left[\ketbra{\Psi^+}{\Psi^+}\sigma_1\right]=1.\label{T1122}\ea 
From Eq.~\eqref{T33} and Eq.~(11) of the main text, we find
$\moy{J_3^{(d_1)}\otimes J_3^{(d_2)}}_{\rho}={j^{(d_1)}j^{(d_2)}}$
which in turn  implies that the state $\rho$ must be a pure state  of
the form $\alpha \ket{j^{(d_1)},j^{(d_2)}}+\beta
\ket{-j^{(d_1)},-j^{(d_2)}}$ with $\abs{\alpha}^2+\abs{\beta}^2=1
$. For such states the other values of
$\moy{\sigma_i}_{\mathcal{M}_{U_1U_2}(\rho)}$ for $i=1$ or $i=2$ are
zero and not 1 or -1, indeed $\moy{J_1^{(d_1)}\otimes J_1^{(d_2)}}_{\rho}=\moy{J_2^{(d_1)}\otimes J_2^{(d_2)}}_{\rho} =0$ for $d_1>2$ or $d_2>2$. The same reasoning can be made for each Bell state.

\section{3-parameter family of  2-qudit states }
\subsection{Preliminaries}
First, we recall the family of $d^2$ maximally entangled 2-qudit states $\ket{\Omega_{kl}} (k,l=0,1,\cdots d-1)$ that  generalizes the four 2-qubit Bell states as introduced in~ Ref.~\cite{Bb2007}:
\ba
\label{Bell}
\ket{\Omega_{kl}}= W_{kl}\otimes\One\ket{\Omega_{00}}~\text{with}~   \ket{\Omega_{00}}=\frac{1}{\sqrt{d}}\sum_{m=0}^{d-1}\ket{m,m},
\ea
where the $d^2$ $W_{kl}$ operators acting on the first qudit are the Weyl operators defined as:
\ba
\hspace{-1em} W_{kl}\ket{m}=w^{k(m-l)}\ket{(m-l)_{\text{mod }d}},~\text{where}~ w=e^{2\pi i/d}.
\ea
These operators verify the following orthogonality relation in the Hilbert-Schmidt norm:
\be
\label{Eq:orth}
\text{Tr}\left[W_{kl}^{\dagger}W_{mn}\right]=d~\delta_{km}\delta_{ln}.
\ee
These 2-qudit Bell states $P_{kl}=\ketbra{\Omega_{kl}}{\Omega_{kl}}$ are locally maximally mixed state, that is, their partial trace gives the maximally mixed state $\frac{\One}{d}$. 

Using the following relation~\cite{Bb2007}: 
\be 
\ketbra{j}{k}=\dfrac{1}{d}\sum_{l=0}^{d-1}w^{lj}W_{j (k-j)},
\ee 
each state $P_{kl}$ can be written in the basis $W_{kl}\otimes W_{mn}$ as follows~\cite{Bb2008,Bb2007}: 
\ba
\label{Eq:P_klinW_kl}
P_{kl}=\dfrac{1}{d^2}\sum_{m,n=0}^{d-1}w^{ml-kn}W_{-m-n}^{\dagger}\otimes W_{m-n}^{\dagger}.
\ea
The operators $J_3,~J_+,~J_-$ can also be written in the Weyl basis:
\ba
\label{Eq:JinW_kl}
J_3&=&\sum_{l=0}^{2j}\eta_z^lW_{l0},\nonumber\\J_+&=&\sum_{l=0}^{2j}\eta_p^lw^{-l}W_{l,-1}~,~J_-=\sum_{l=0}^{2j}\eta_p^lW_{l,1},
\ea
where we have defined
\ba
\eta_z^l&=&\sum_{m=0}^{2j}\frac{m-j}{2j+1}w^{-ml},\nonumber\\ \eta_p^l&=&\sum_{m=0}^{2j}\frac{\sqrt{j(j+1)-(m-j)(m-j+1)}}{2j+1}w^{-ml}.\nonumber
\ea

From Eqs.~\eqref{Eq:P_klinW_kl},~\eqref{Eq:JinW_kl}, and~\eqref{Eq:orth}, we get
\begin{eqnarray}
\moy{J_3\otimes J_3}_{\ket{\Omega_{kl}}}&=&\text{Tr}\left[P_{kl}J_3\otimes J_3\right]=\sum_{m=0}^{2j}\eta_z^m\eta_z^{-m}w^{ml},\nonumber\\
\moy{J_+\otimes J_+}_{\ket{\Omega_{kl}}}&=&\text{Tr}\left[P_{kl}J_+\otimes J_+\right]=w^{-k}\sum_{m=0}^{2j}\eta_p^m\eta_p^{-m}w^{ml},\nonumber\\
\moy{J_+\otimes J_-}_{\ket{\Omega_{kl}}}&=&\text{Tr}\left[P_{kl}J_+\otimes J_-\right]=0.\nonumber
\end{eqnarray} 
Now, we have the tools that enable us to study the three-parameter family of 2-qudit states with $d=2j+1$ defined in the main text:
\ba
\label{rhod_abc}
\rho_{\alpha,\beta,\gamma}&=&\dfrac{1-\alpha-\beta-\gamma}{(2j+1)^2}\One+\alpha P_{00}\nonumber \\&+&\dfrac{\beta}{2j}\sum_{i=1}^{2j}P_{i0}+\dfrac{\gamma}{2j+1}\sum_{i=0}^{2j}P_{i1}.
\ea  

For these states we have
\ba
\moy{J_3\otimes J_3}_{\rho_{\alpha,\beta,\gamma}}&=& \dfrac{j(j+1)\left(\alpha+\beta\right)+j(j-2)\gamma}{3},\nonumber\\
\moy{J_1\otimes J_1}_{\rho_{\alpha,\beta,\gamma}}&=& \dfrac{j(j+1)\left(\alpha-\beta/2j\right)}{3},\nonumber\\
\moy{J_2\otimes J_2}_{\rho_{\alpha,\beta,\gamma}}&=& -\dfrac{j(j+1)\left(\alpha-\beta/2j\right)}{3},\nonumber
\ea
so that the criterion \begin{eqnarray}
 \abs{\moy{J_1^{(d_1)}\otimes J_1^{(d_2)}}_{\rho}}&+&\abs{\moy{J_2^{(d_1)}\otimes J_2^{(d_2)}}_{\rho}}\nonumber\\&+&\abs{\moy{J_3^{(d_1)}\otimes J_3^{(d_2)}}_{\rho}}\leq  {j^{(d_1)}j^{(d_2)}} \nonumber
 \end{eqnarray} introduced in the main text for separable states $\rho_{\alpha,\beta,\gamma}$ can be brought to the following form:
\begin{equation}
\label{critabc}
\hspace{-1em}\dfrac{\abs{ (j+1) (\alpha+\beta) +(j-2) \gamma }+(j+1)  \abs{2 \alpha-\beta/j} }{3j}\leq 1.
\end{equation}

\subsection{Positivity and partial transpose of~$\rho_{\alpha,\beta,\gamma}$}
From the definition in Eq.~\eqref{rhod_abc},
$\rho_{\alpha,\beta,\gamma}$ is already in its diagonal form and we
can easily identify its eigenvalues and eigenvectors. The eigenvalues
are:$\dfrac{1-\alpha-\beta-\gamma}{(2j+1)^2}+\dfrac{\gamma}{2j+1}$
with degeneracy $2j+1$,
~$\hspace{-0.4em}\dfrac{1-\alpha-\beta-\gamma}{(2j+1)^2}+\dfrac{\beta}{2j}$
with degeneracy $2j$,
$\dfrac{1-\alpha-\beta-\gamma}{(2j+1)^2}~+~\alpha$ with degeneracy
$1$, and $\dfrac{1-\alpha-\beta-\gamma}{(2j+1)^2}$ with degeneracy
$(2j+1)(2j-1)$. To ensure the positivity of
$\rho_{\alpha,\beta,\gamma}$, each of these eigenvalues must be
positive, hence we obtain the positivity conditions stated in the main article:
\begin{eqnarray}
 & \dfrac{1-\alpha-\beta-\gamma }{(2j+1)^2} \geq 0  ,&  \alpha+\dfrac{1-\alpha-\beta-\gamma }{(2j+1)^2}  \geq 0, \nonumber \\ 
& \dfrac{\beta}{2j}+\dfrac{1-\alpha-\beta-\gamma }{(2j+1)^2} \geq 0 ,&  \quad \dfrac{1-\alpha-\beta+2j\gamma }{(2j+1)^2} \geq 0. \nonumber
\end{eqnarray}
As for the partial transpose (PT) of $\rho_{\alpha,\beta,\gamma}$, it
was shown in Ref.~\citep{Bb2008} that for a state of the form $\rho=\sum_{k,l=0}^{2j}C_{kl}P_{kl}$, its partial transpose can be written as:
\be
\rho^{T_B}={\bigoplus}_{m=0}^{2j}B_m \quad: B_m^{\dagger}=B_m,
\ee
where, the matrix elements of $B_m$ are defined as~\cite{Bb2008}: 
\be
\label{Eq:pt}
\bra{s}B_m\ket{t}=\dfrac{1}{2j+1}\sum_{k=0}^{2j}C_{s,s+t-m}w^{k(s-t)}: s,t=0,\cdots,2j.
\ee
It was also shown~\citep{Bb2008} that for  integer values of $j$, all
matrices $B_m$ are unitarily equivalent, while for $j$ half-integer,
there are two classes of unitarily equivalent matrices, the class of
$B_m$ for even $m$ and the class for odd $m$.

In what follows, we will restrict ourselves to the case of $j$ integer and we will calculate the elements of the matrix $B_0$. The case of half-integer $j$ can be easily obtained based on the calculation of the matrix $B_0$.

From Eqs.~\eqref{rhod_abc} and~\eqref{Eq:pt}, we find that the matrix $B_0$ has the following form:\vspace{-1em}

\be
B_0=
\begin{pmatrix}
b_{00} &0       & \cdots & \cdots &\cdots & \hspace{-0.5cm} 0 \\

0      & b_{11} &0        & \cdots &0     &\hspace{-0.5cm}\kappa \\
\vdots & 0      &\ddots   &        &\text{\reflectbox{$\ddots$}}&\hspace{-0.5cm}0  \\

\vdots & \vdots &         &   \ddots     &                             &\hspace{-0.5 cm}\vdots\\

0      & 0      &    \text{\reflectbox{$\ddots$}}      &       &      \hspace{-0.5 cm}       b_{2j-1,2j-1}       &\hspace{-0.5cm} 0 \\
0      &\kappa & 0         & \cdots &        0                   & b_{2j,2j}    
\end{pmatrix},
 \ee
 with
\ba
&b_{00}&=\bra{0}B_0\ket{0}=\dfrac{1-\alpha-\beta-\gamma}{(2j+1)^2}+\dfrac{\alpha+\beta}{2j+1},\nonumber\\
&b_{j+1,j+1}&=\bra{j+1}B_0\ket{j+1}=\dfrac{1-\alpha-\beta-\gamma}{(2j+1)^2}+\dfrac{\gamma}{2j+1},\nonumber\\
&b_{k,k}&=\bra{k}B_0\ket{k}=\dfrac{1-\alpha-\beta-\gamma}{(2j+1)^2} :k\in\lbrace0,\cdots,2j\rbrace\setminus\lbrace0,j+1\rbrace, \nonumber\\
&\kappa &=\bra{s}B_0\ket{t}=\dfrac{\alpha-\beta/2j}{2j+1} \hspace{2.5em} : s+t=0~\text{and}~s\neq t ,\nonumber\\
&\bra{s}B_0\ket{t} &= 0 \hspace{11em}: s\neq t~\text{and} s+t\neq 0.\nonumber
\ea
The characteristic polynomial of $B_0$ can be easily computed which
allows to calculate the eigenvalues of $B_0$. As
$\rho_{\alpha,\beta,\gamma}^{T_B}$  is the direct sum of matrices
$B_m$ that
are unitarily equivalent to $B_0$, then the eigenvalues of $B_0$ are
also those of $\rho_{\alpha,\beta,\gamma}^{T_B}$ .
We distinguish 2 cases:
for $j=1$ and $j>1$. For $j=1$, the eigenvalues are:
\ba
&&\frac{1+2 j (\alpha +\beta )-\gamma }{(1+2 j)^2},\nonumber\\&&\frac{1}{2} \left(\frac{\gamma }{1+2 j}-\frac{2 (-1+\alpha +\beta +\gamma )}{(1+2 j)^2}-\sqrt{\frac{(-2 j \alpha +\beta )^2+j^2 \gamma ^2}{j^2 (1+2 j)^2}}\right),\nonumber \\&& \frac{1}{2} \left(\frac{\gamma }{1+2 j}-\frac{2 (-1+\alpha +\beta +\gamma )}{(1+2 j)^2}+\sqrt{\frac{(-2 j \alpha +\beta )^2+j^2 \gamma ^2}{j^2 (1+2 j)^2}}\right)\nonumber
\ea
with degeneracy $3$  each. Whereas for $j>1$, the eigenvalues are:
\ba
&&\frac{1+2 j (\alpha +\beta )-\gamma }{(1+2 j)^2},\nonumber\\&&\frac{-\beta -2 j (-1-2 j \alpha +2 \beta +\gamma )}{2 j (1+2 j)^2},\frac{\beta -2 j (-1+2 (1+j) \alpha +\gamma )}{2 j (1+2 j)^2},\nonumber \\ &&\frac{1}{2} \left(\frac{\gamma }{1+2 j}-\frac{2 (-1+\alpha +\beta +\gamma )}{(1+2 j)^2}-\sqrt{\frac{(-2 j \alpha +\beta )^2+j^2 \gamma ^2}{j^2 (1+2 j)^2}}\right),\nonumber \\&& \frac{1}{2} \left(\frac{\gamma }{1+2 j}-\frac{2 (-1+\alpha +\beta +\gamma )}{(1+2 j)^2}+\sqrt{\frac{(-2 j \alpha +\beta )^2+j^2 \gamma ^2}{j^2 (1+2 j)^2}}\right)\nonumber
\ea
with degeneracy $2j+1,~(2j+1)(j-1),~(2j+1)(j-1)$, $(2j+1)$, and
$(2j+1)$ correspondingly. Thus by imposing positivity on the above
eigenvalues, we get the set of conditions for the state
$\rho_{\alpha,\beta,\gamma}$ to be PPT. This is how we compute the
blue region on  figure~1 (a) and (b) in the main text.

\end{document}